\documentclass[amsmath,aps,prb,twocolumn]{revtex4}
\usepackage{epsfig}

\begin{document}
\title{Effect of disorder and electron-phonon interaction on
interlayer tunnelling current in quantum Hall bilayer.}
\author{O.G.C.~Ros$^1$ and D.K.K.~Lee$^2$}
\affiliation{$^1$Instituto Pluridisìplinar, Universidad Complutense Paseo Juan XXIII, 1,
28040 Madrid, Spain\\
$^2$Blackett Laboratory, Prince Consort Rd, Imperial College London, London SW7 2AZ,
United Kingdom.}
\date{\today}

\newcommand{\bq}{\mathbf{q}}
\newcommand{\bQ}{\mathbf{Q}}
\newcommand{\bk}{\mathbf{k}}
\newcommand{\dsas}{\Delta_\perp}
\newcommand{\inversemicron}{$\mu$m$^{-1}$}

\begin{abstract}
We study the transport properties of the quantum Hall bilayers systems
looking closely at the effect that disorder and electron-phonon
interaction have on the interlayer tunnelling current in the presence
of an in-plane magnetic field $B_\parallel$.
We find that it is important to take into account the effect of
disorder and electron-phonon interactions in order to predict a finite
current at a finite voltage when an in-plane magnetic field is
present. We find a broadened resonant feature in the tunnelling
current as a function of bias voltage, in qualitative agreement with
experiments. We also find the broadening due
to electron-phonon coupling has a non-monotonic dependence on
$B_\parallel$, related to the geometry of the double quantum well. We
also compare this with the broadening effect due to spatial
fluctuations of the tunnelling amplitude. We conclude that such static disorder
provides only very weak broadening of the resonant feature in the experimental range.

\end{abstract}

\maketitle

\section{Introduction}\label{int}
Over the last fifteen years quantum Hall bilayer systems (QHB) have
been extensively studied since they are one of the few systems that
show macroscopic evidence of quantum coherence. The richness of the
physics of the QHB has attracted the attention of both theoretical
\cite{girvin,moon,wen,balents,stern,jacklee} and
experimental studies \cite{eis,mur,kellogg,spielman,spielman2}.
This has led to rather rapid progress in the area. The bilayer
consists of two parallel two-dimensional electron layers in a double
quantum well closely separated by a distance $d$ and subjected to a
magnetic field perpendicular to the plane of the layers $B_\perp$.
In this paper, we will focus on the case when  
each layer is a half-filled Landau level: filling factor $\nu_1=\nu_2=1/2$. If the
separation between layers is large they behave as two independent
Fermi liquids and no quantum Hall effect is observed. When the
distance between the layers becomes comparable with the magnetic
length ($d\sim\ell_B)$ the system undergoes a phase transition from
a compressible state at large $d\gg \ell_B$ to an incompressible state
small $d\ll \ell_B$. In the incompressible state, the system
as a whole exhibits the $\nu=1$ quantum Hall effect even when interlayer
tunnelling is negligible. This transition is driven by Coulomb
interactions between the layers. The ground state in the quantum
Hall regime is believed\cite{wen} to have a broken U(1) symmetry which leads
to spontaneous interlayer phase coherence.
This ground state can be described as a pseudospin ferromagnet
\cite{moon} using a pseudospin picture equating electrons in the
upper (lower) layer with pseudospin ``up'' (``down'')
\cite{moon,girvinleshouches}, or as an excitonic superfluid\cite{balents,fertig}.

A series of remarkable experiments have probed the existence of this
coherent phase. They show evidence for interlayer coherence with a
linearly dispersing Goldstone mode \cite{spielman2} and counterflow
superfluidity\cite{kellogg,tutuc,tiemann2} and drag Hall
voltage\cite{tiemann}.  One piece of the experimental evidence of
interlayer coherence is a sharp peak in the tunnelling current for
small bias (between 10 and 100$\mu$V) and low temperature. Another
characteristic of the QHB that indicates the existence of phase
coherence between layers is the sensitivity of the QHB to the presence
of an in-plane magnetic field. The sharp peak in the tunnelling
current at small bias is suppressed when a magnetic field is applied
parallel to the plane of the layers. At the same time, experiments
show a 'dispersive' feature in the tunnelling current at a voltage
which evolves linearly with the in-plane field $B_\parallel$.  This
has been interpreted as the excitation of the Goldstone mode of the
excitonic superfluid at wavevector $Q$ and energy $eV$ given by
\begin{equation}\label{Qdefine:eq}
Q= \frac{2\pi B_\parallel d}{\phi_0}, \qquad eV=\hbar vQ
\end{equation}
where $v$ is the velocity of the collective mode and $\phi_0 = hc/e$ is the flux quantum
\cite{spielman2}.

The tunnelling current can be computed using as a perturbation the
interlayer hopping matrix element $t_\perp$. For a homogeneous system
in the absence of an in-plane field, Jack \emph{et al}\cite{jacklee}
found a tunnelling current proportional to $1/V$ at zero temperature,
consistent with the observation of a region of negative differential
tunnelling conductance at low temperature. However, the same
calculation gives a delta function at $eV = vQ$ in the tunnelling
current in the presence of an in-plane field. Although the position of
this feature is consistent with experiments, the experimental data
does not exhibit a sharp feature even at the lowest temperatures.  In
this paper, we will explore sources for a finite linewidth of this
feature. We find that two mechanisms should be dominant at low
temperatures: tunnelling disorder and electron-phonon coupling. We will not
discuss the role of vortices which could be nucleated at zero temperature
by strong charge disorder or by thermal activation.

This paper is organised as follows. In the next section, we review the
methodology for computing the tunnelling current as originally used by
Jack \emph{et al}\cite{jacklee}. In Section \ref{pei}, we discuss the
effect of electron-phonon interactions on the tunnelling current. We
will calculate how the magnitude and the width of the dispersive feature in
the tunnelling current is affected by the electron-phonon coupling. In
Section \ref{dis}, we investigate the effect that an inhomogenous
tunnelling amplitude over the sample has on the tunnelling
current. We present the conclusions of the paper in Section \ref{sumphon:sec}.
\section{Theoretical discussion}\label{gdphon}
We will adopt the pseudospin picture of the quantum Hall bilayer,
labelling single-particle states in the upper layer as $S^z=+1/2$ and
states in the lower layer as $S^z=-1/2$. We will review this framework
in this section.  In this picture, the system is a pseudospin
ferromagnet with an easy-plane anisotropy. Furthermore, we will work
with the large-$S$ generalisation of this model on a lattice which
corresponds to coarse-graining the system by treating ferromagnetic
patches containing $S$ electrons as lattice sites containing a large
spin $S$.  The Hamiltonian can be written as:
\begin{equation}
H_0 = -\frac{J}{2}\sum_{\langle ij\rangle}(S^+_i  S^-_j + \mathrm{h.c.})
+\frac{D}{2}\sum_i (S^z_i)^2
\label{H0:eq}
\end{equation}
where $S^+_i$ and $S^-_i$ are the pseudospin raising and lowering
operators on site $i$ of a square lattice,
$J$ is the exchange interaction, $D$ represents a local capacitative
energy for charge imbalance. In the presence of tunnelling across the bilayer,
the Hamiltonian becomes:
\begin{equation}
H = H_0 - \dsas \sum_i (e^{iQx_i} S^+_i + e^{-iQx_i} S^-_i)
\label{Hclean:eq}
\end{equation}
where $\dsas$ is the interlayer
tunnelling matrix element in the absence of in-plane field and
$x_i$ is the $x$-coordinate
of the spin $i$.
We have chosen the gauge such that, in the
presence of an in-plane magnetic field $B_\parallel$, the tunnelling
matrix acquires a phase that varies spatially in the $x$-direction with periodicity $2\pi/Q$
with $Q$ as defined in (\ref{Qdefine:eq}).

The large-$S$ treatment of this model corresponds to taking $S$ to a
large value while keeping $JS$ and $DS$ constant so that the three
energy scales for exchange, interaction and tunnelling scale in the
same way with $S$. We note that $DS > JS \gg \dsas$ in the bilayer
system.

The pseudospin can be written in terms of phase and $S^z$
operators as
\begin{equation}\label{Sz}
S_i^+=(S-S^z_i)^{1/2}e^{-i\phi_i}(S+S^z_i)^{1/2}
\end{equation}
Semiclassically, $\phi_i$ gives the azimuthal angle of the pseudospin
projected onto the $xy$-plane in spin space.  They are canonical
conjugate variables: $[S^z_i,\phi_j]=i\delta_{ij}$.  We can take the
continuum limit and integrate out the $S^z$ (charge imbalance)
fluctuations to arrive at a phase-only action\cite{kun}:
\begin{equation}\label{phaseonly:lag}
{\cal S}=2S\rho_s\!\!\int\!\! d^2r dt
\left[\frac{1}{2v^2}(\partial_t\phi)^2\! 
-\frac{1}{2}|\nabla\phi|^2+\frac{1}{\xi^2}\cos(\phi-Qx) \right]
\end{equation}
where $v^2=DSJS\ell_B^2/\hbar^2$, $\rho_s=JS$ is the spin stiffness and
$\xi = (4\pi\rho_s/\dsas)^{1/2}\ell_B$ is the Josephson length
which gives the length scale over which
counterflow currents decay due to tunnelling across the bilayer.
If we further include a bias of $V$ across the bilayer,
the only change to the action is that the cosine term 
in the action becomes\cite{balents,jacklee} $\cos(\phi -Qx - eVt/\hbar)$. 
In this model, all the dynamics of the system
depends on the phase $\phi$. Spatial gradients in the phase correspond to
counterflow in the two layers.

In the absence of tunnelling, this Lagrangian represents an easy-plane
ferromagnet with spontaneously broken symmetry, \emph{i.e.} a
spatially uniform phase.  In the presence of an in-plane magnetic
field, the ground state develops spatial variations in the phase
field\cite{bak,hanna}.  At small $Q$, the phase field increases
linearly with the Aharonov-Bohm phase: $\phi \sim Qx$. At large $Q$,
the phase field cannot follow the Aharonov-Bohm phase and has only
small oscillations: $\phi \sim \sin(Qx)/Q^2\xi^2$ where $\xi$ is the
Josephson length. The transition occurs at $Q\xi \sim$ O(1). The
Josephson length is estimated to be of the order of
microns. Experimental values\cite{spielman2} of the in-plane field
give $Q\xi > 10$. Therefore, the spatial fluctuations of the
ground-state phase field are small. In our following calculation, we
will consider quantum fluctuations as perturbations around the uniform
$\phi=0$ state.

Our calculation of the quantum fluctuations in
the system starts by the system in the absence of tunnelling. Then we
will introduce interlayer tunnelling in perturbation theory. In the absence of
tunnelling, the $\dsas=0$ Hamiltonian (\ref{H0:eq}) can be diagonalized
\begin{eqnarray}
H_0&=&\sum_{\bq\neq0}\varepsilon_\bq\alpha^\dag_\bq\alpha_\bq\,,\nonumber\\
\varepsilon_\bq &=& [JS q^2 l_B^2 ( DS + q^2 l_B^2)]^{1/2},\nonumber\\
\alpha^\dag_{\bq}&=&\sqrt{\frac{S}{2}}\left[ (u_\bq + v_\bq)
\frac{S^z_{-\bq}}{S}+i(u_\bq - v_\bq)\phi_{-\bq}\right]
\label{alpha}
\end{eqnarray}
where the coherence factors are $(u_\bq + v_\bq)^2=(u_\bq -
v_\bq)^{-2}= (1 + DS/JS q^2\ell_B^2)^{1/2}$.  This means that the
elementary excitations in the pseudospin lattice system, created by
the operator $\alpha^\dag_\bq$ are long-lived pseudospin waves with
energy $\varepsilon_\bq$ at wavevector $\bq$. At long wavelengths, 
the pseudospin waves have a linear dispersion, $\varepsilon_\bq \simeq \hbar vq$,
with pseudospin wave velocity $v \sim (JS DS)^{1/2} l_B/\hbar$.

The operators $S^z$, $e^{\pm i\phi}$ and $S^\pm$ all involve the creation
and annihilation of pseudospin waves.  Most significantly, interlayer
tunnelling, $S^\pm$, causes decay of the pseudospin waves at all wavelengths.
This can be seen by examining the $S^+_i$ operator that represents
electron tunnelling in the pseudospin language. From equation
(\ref{Sz}), we see that it creates perturbations in the phase $\phi_i$
and $S^z_i$. This involves the creation and annihilation of pseudospin
waves (see equation (\ref{alpha})).  This is a consequence of the fact
that tunnelling breaks the global U(1) phase invariance of the system
so that Goldstone's theorem no longer protects the long-wavelength
pseudospin waves from decay.

Let us consider now the interlayer tunnelling
current in the presence of an interlayer bias $V$.  The tunnelling
current at site $i$ of the lattice is given by the operator $i\dsas
(e^{-i(Qx_i-eVt/\hbar)}S_i^- - e^{i(Qx_i-eVt/\hbar}S_i^+)/2\hbar$.
Therefore, in the continuum limit, the expectation value of the
interlayer tunnelling current is\cite{balents,jacklee}:
\begin{equation}
I(t) = \frac{e\dsas}{2\pi\hbar l_B^2}\int\!d^2r
\left\langle\sin\left(\phi-Qx-\frac{eVt}{\hbar}\right)
\right\rangle
\label{currentdef:eq}
\end{equation}
where the expectation value is taken with respect to the full
Hamiltonian $H$ as defined by (\ref{Hclean:eq}). In a perturbative treatment of the
interlayer tunnelling in the Hamiltonian $H$, we can
treat the tunnelling term in first-order perturbation theory.  This
gives\cite{balents,jacklee} a dc current proportional to $\dsas^2$:
\begin{equation}\label{phoncurr:eq}
I_0=-Se\left(\frac{\dsas L}{4\pi\hbar\ell^2_B}\right)^2\mathrm{Re}
\int\!\! d^2\!r\!\int_0^\infty\!\!\!\!\!\!d\tau
C(\mathbf{r},\tau)e^{-i(\bQ\cdot\mathbf{r}+eV\tau/\hbar)}
\end{equation}
where $\bQ=(Q,0)$, $C(\mathbf{r},\tau) = (\langle
Te^{i\phi(\mathbf{r},\tau)}e^{-i\phi(0,0)}\rangle_0-\mathrm{h.c.})$
evaluated at zero tunnelling with Hamiltonian $H_0$. We can see that
we are calculating the response of the system at wavevector $Q$ and frequency $eV/\hbar$.
It can be shown that
\begin{equation}\label{to}
\ln \langle Te^{i\phi(\mathbf{r},\tau)}e^{-i\phi(\mathbf{r}',0)}\rangle_0=
\frac{-i\hbar}{2S\rho_s}\!\!\int\!\!\!\frac{d^2qd\omega}{(2\pi)^3}
|J_{\mathbf{q},\omega}(\mathbf{r},\mathbf{r}',\tau)|^2 G^{(0)}_{\bq,\omega}
\end{equation}
where
$J_{\mathbf{q},\omega}(\mathbf{r},\mathbf{r}',\tau)=
e^{i(\mathbf{q}\cdot\mathbf{r}-\omega\tau)}-e^{i\mathbf{q}\cdot
  \mathbf{r}'}$ and $G^{(0)}_{\bq,\omega}= (\omega^2/v^2-q^2)^{-1}$ is
the propagator for the phase fluctuations at zero tunnelling.

Jack \emph{et
  al}\cite{jacklee} analysed the tunnelling current at zero in-plane
field.  They showed that the perturbative calculation above can be
understood in terms of the generation of finite-momentum pseudospin
waves \emph{via} the decay of the $q=0$ mode. Technically, this is an
interpretation of (\ref{phoncurr:eq}) as a Taylor expansion of the
exponential in the definition of $C(\mathbf{r},t)$. Each term in the
Taylor expansion involving $2n$ $\phi$-fields represents the
generation of $n$ pseudospin waves. In the absence of an in-plane
field, there is no decay of the $q=0$ mode to a single pseudospin wave
with non-zero wavevector because of momentum conservation. The most
important decay channel is then the generation of a pair of pseudospin
waves with equal and opposite momenta. However, in the presence of an
in-plane magnetic field $B_\parallel$, the vector potential provides a
momentum of $Q$, as given by equation (\ref{Qdefine:eq}), to the
pseudospin system, as can be seen in the Hamiltonian
(\ref{Hclean:eq}). The generation of a single pseudospin wave at
wavevector $Q$ is now possible and this is the leading contribution in
orders of $1/S$. (In the perturbative formulation, this can be seen
mathematically in (\ref{phoncurr:eq}) which tells us that we need to
calculate the Fourier component of $C$ at wavevector $\mathbf{Q}$.)
For a homogeneous system, this calculation gives a current which is a
delta function at $eV/\hbar=vQ$:
\begin{equation}
	I_0 = \frac{e\dsas^2 L^2}{\hbar l_B^2}\frac{DS}{8eV}\delta(eV-\hbar vQ)
\label{currdelta:eq}
\end{equation}
In this work, we investigate possible sources of line broadening for
this peak at non-zero in-plane magnetic field.  We will focus on
effects which do not vanish at zero temperature. In order to obtain a
finite linewidth, we find that the pseudospin waves need elastic or
inelastic scattering. We will discuss elastic scattering due to disorder in section \ref{dis}.
In section \ref{pei}, we will study inelastic scattering. This can
arise from the generation of photons or phonons.  The two mechanisms
are similar. However, we will see that the energy of photons involved
in this process will be much higher than the energy of the pseudospin
waves. This means that the process can only be virtual.  It can at
most alter the dispersion relation of the pseudospin waves but cannot
cause decay.  Therefore, we will focus on the electron-phonon
interaction in this work.

\section{phonon generation}\label{pei}
In this section, we introduce interactions between phonons and
pseudospin waves and study how the tunnelling current between layers
is affected by the introduction of these interactions.

When electrons tunnel across the bilayer, the electron density
changes, perturbing the core ions on the AlGaAs of the tunnelling
barrier and thus creates phonons in the three-dimensional system in
which the quantum well is embedded.  For the range of values of the
bias voltage used in the experiments, the most important interaction
between ions in the host material and the tunnelling electrons is the
deformation potential interaction \cite{mahan,cardona} with the
acoustic phonons. We neglect optical phonons because they are
at energies high compared to the electron energy at the experimental
range of bias voltage.

The phonon Hamiltonian is given by: 
\begin{equation}\label{Hphonon}
H_{\rm phonon} =
\sum_{\mathbf{k}} \hbar\omega_{\bk} \left(a^\dagger_\mathbf{k} a_\mathbf{k} +
\frac{1}{2}\right)
\end{equation}
where $\mathbf{k}=(\mathbf{q},k_z)$ is a three-dimensional wavevector,
$\mathbf{q}=(q_x,q_y)$, $\omega_\mathbf{k}$ is
the phonon frequency spectrum and
$a_\mathbf{k}$ and $a_\mathbf{k}^\dagger$ are the phonon annihilation and creation operators.
We are discussing physics at wavelengths long
compared to the lattice spacing of the substrate.  So, we will use a
simple linear dispersion for the acoustic phonon: $\omega_\mathbf{k}= s|\bk|$.
The electron-ion interaction takes the form:
\begin{equation}\label{def}
H_\mathrm{e-ion}=D_\mathrm{ef}\sum_\mathbf{k}
\left(\frac{\hbar}{2\rho_\mathrm{ion}\nu\omega_\mathbf{k}}\right)^{1/2} |\mathbf{k}|
\rho(\mathbf{k})(a_\mathbf{k}+a_{-\mathbf{k}}^\dagger)
\end{equation}
where $\rho_\mathrm{ion}$ is the ion mass
density, $\nu$ is the volume of the three-dimensional solid
and $\rho(\mathbf{k})$ is the three-dimensional electron density.

The electron density perturbation caused by tunnelling involves charge imbalance across the bilayer.
It is convenient to express the density perturbation at position $\mathbf{R}=(\mathbf{r},z)$
in terms of the $z$-component of the pseudospin:
\begin{equation}
\rho(\mathbf{R})=\frac{S}{4\pi\ell_B^2}\left[\rho_{\uparrow}(z)+\rho_{\downarrow}(z)
  +\frac{S^z(\mathbf{r})}{S}\left(\rho_\uparrow(z)-\rho_\downarrow(z)\right)\right]
\end{equation}
where $\rho(\mathbf{R})$ is the three-dimensional electron density and
$\rho_{\uparrow(\downarrow)}(z)$ is the one-dimensional density
profile in the $z$-direction for electrons in the upper (lower) layer,
normalized to $\int\rho_{\uparrow(\downarrow)}(z)dz =1$. We have scaled
the density by $S$ in the spirit of the coarse-graining idea of the
large-$S$ generalisation of this model, as discussed at the start of
section \ref{gdphon}. It is easy to check that the above expression  produce
the expected three-dimensional density profile for the cases of 
to a balanced bilayer $S_z=0$ and completely imbalanced bilayer $S_z = \pm S$.

\begin{figure}
\begin{center}
  \epsfig{file=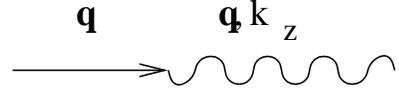, width=.6\columnwidth}
\caption{\small{Diagrammatic representation of the decay of one
pseudospin wave with momentum $\mathbf{q}$
into one phonon with momentum $\mathbf{q},k_z$.}}\label{phonondecay}
\end{center}
\end{figure}

For simplicity, we approximate the electron wavefunction
$\psi(z)$ in the lowest subband of the single quantum well centered at
$z=0$ by a Gaussian wavepacket:
\begin{equation}\label{gaussian}
\psi(z)\simeq\frac{\pi^{1/4}}{w^{1/2}}e^{-z^2\pi^2/2w^2}
\end{equation}
where $w$ is the width of the well. (A factor of $\pi^2$
has been inserted into the exponential so that the form matches the
amplitude and width of the actual wavefunction in the square well.)
Using this approximation, the electron density in the Fourier
representation is given by
\begin{equation}\label{den}
\rho(\mathbf{k})=\frac{i}{2\pi\ell_B^2}\sin\left(\frac{k_zd}{2}\right)e^{-k_z^2w^2/4\pi^2}
S^z_\bq + \ldots
\end{equation}
where $d$ is the separation of the centres of the two quantum wells in
the $z$-direction. We have dropped terms
that do not involve the pseudospin degrees of freedom, \emph{i.e.}
terms that do not change as a result of a tunnelling event that
creates a charge imbalance across the bilayer. 
Note that the part proportional to $S^z$
vanishes as $k_z\to 0$ because this part represents charge imbalance
and does not contribute to a uniform change in density in the $z$-direction.
In fact, the form of the $k_z$-dependence
in $\rho(\mathbf{k})$ reflects the geometry of
the double quantum well. There are two contributions to the
geometrical form factors in this expression, analogous to the
Fraunhofer diffraction pattern from a double slit in optics. The
sinusoidal dependence on $k_z d/2$ comes from the convolution of the
two density profiles centered at $\pm d/2$
and the Gaussian comes from the electron density profile in the
$z$-direction. 
From equations (\ref{Sz}), (\ref{alpha}) and (\ref{den}), we see that
the electron-phonon interaction in equation (\ref{def}) includes the
decay of one pseudospin wave with momentum $\mathbf{q}$ into one
phonon with momentum $(\mathbf{q},k_z)$ (or \emph{vice versa}).  This
is depicted in Figure \ref{phonondecay}. There is no conservation of
momentum in the $z$-direction because translational symmetry is broken
by the presence of the double quantum well.  This decay of the
pseudospin wave into a phonon can only occur as a real process if
energy is conserved:
\begin{equation}
vq=s\sqrt{q^2+k_z^2}.
\end{equation}
This requires the pseudospin wave velocity to be higher than the
phonon velocity: $v>s$. This is the case we consider here. We estimate $v\sim
1.4\times 10^4$ ms$^{-1}$ from the data of Spielman \emph{et
  al.}\cite{spielman2} and $s \simeq 0.1 v$ for the sound
velocity in the heterostructure\cite{cardona}.  
On the other hand, if the phonon velocity $s$ was
greater than the pseudospin wave velocity $v$, then the process where
a phonon of greater speed (or indeed a photon) can be emitted and
re-absorbed is only a virtual process. This will alter the pseudospin
wave spectrum but does not contribute to the tunnelling current.

We will now investigate more quantitatively how the pseudospin-phonon
interaction affect the system.  We proceed by integrating out the
phonons to obtain an effective action. This can be performed because
the electron-phonon coupling is linear in the phonon field.  We obtain
a retarded interaction for $S^z$, the charge imbalance across the
layers:
\begin{equation}\label{so}
{\cal S}_{\rm ep}=\frac{2Sv^2}{8\pi \ell_B^4\rho_s}\!
\int\!\!\frac{d^2q\,d\omega}{(2\pi)^3}
\lambda_D\Pi_{\mathbf{q},\omega}
|S^z_{\bq,\omega}|^2
\end{equation}
where
\begin{equation}
\Pi_{\mathbf{q},\omega}=\int
dk_z\frac{\sin^2(k_zd/2)(\mathbf{q}^2+k_z^2)e^{-2k_z^2w^2/4\pi^2}
}{\omega^2-\omega_{\mathbf{k}}^2}
\end{equation}
and
$\lambda_D=\left(D_\mathrm{ef}/\hbar\right)^2(\rho_s/\rho_\mathrm{ion}Sv^2$).
This adds to the on-site instantaneous interaction $D$. Integrating
out $S^z$, we obtain a modified phase action at zero tunnelling:
\begin{equation}
{\cal S}_0=\frac{S\rho_s}{(2\pi)^3}\!\int\!\!d^2q\,d\omega
\left(-\mathbf{q}^2+
\frac{\omega^2/v^2}{1+\lambda_D\Pi_{\mathbf{q},\omega}}\right)
|\phi_{\bq,\omega}|^2
\end{equation}

The decay of the pseudospin wave into a phonon is encoded in the
imaginary part of $\Pi_{\mathbf{q},\omega}$. It is non-zero when
$\omega \geq sq$:
\begin{equation}\label{imsigma:eq}
\mathrm{Im}\Pi_{\bq,\omega}=
-\frac{\omega^2}{s^3}
\frac{\sin^2(d\sqrt{\omega^2-s^2q^2}/2s)e^{-2w^2(\omega^2-s^2q^2)/4\pi^2
    s^2}}{\sqrt{\omega^2-s^2q^2}}
\end{equation}
For $\omega < sq$, the
imaginary part of $\Pi_{\mathbf{q},\omega}$ is zero.  The real part of
$\Pi_{\mathbf{q},\omega}$ will shift the position of the peak of the
tunnelling current with respect to the voltage.

The leading term on the expansion of (\ref{to}) in powers of $1/S$
corresponds to the decay of one $q=0$ mode into one finite-momentum
spin wave at wavevector $\mathbf{Q}=(Q,0)$.
The contribution to the current from this term is
\begin{eqnarray}\label{currentG}
I=Se\left(\frac{\dsas L}{4\pi\hbar\ell^2_B}\right)^2&&
\!\!\!\!\!\!\!\!\!\!\!\mathrm{Re}
\!\!\int_0^\infty\!\!\! d\tau\!\! \int_{-\infty}^\infty\!\!\!\!\!d\omega
e^{ieV\tau/\hbar} \cos(\omega\tau)\times\nonumber\\
&&\left(-\frac{i\hbar}{S\rho_s}G_{\mathbf{Q},\omega}-\mathrm{h.c.}\right)
\end{eqnarray}
with
\begin{figure}
\begin{center}
  \epsfig{file=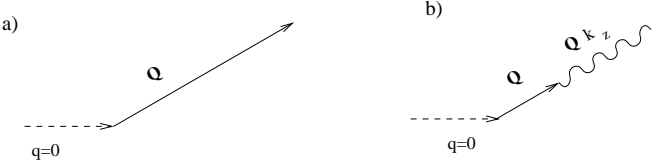, width=\columnwidth}
\caption{\small{Diagrammatic representation of: a) the decay of one
    $q=0$ pseudospin wave into a pseudospin wave with finite momentum
    $\mathbf{Q}$.  b) the decay of one pseudospin wave with momentum
    $q=0$ into a finite-$\mathbf{Q}$ pseudospin wave which in turn
    decays into a phonon.}}\label{qzerophonondecay}\label{dec}
\end{center}
\end{figure}
\begin{equation}
G_{\mathbf{Q},\omega}^{-1}=
[(1+\lambda_D\Pi_{\mathbf{Q},\omega})^{-1}\omega^2/v^2-\mathbf{Q}^2].
\end{equation}
The main difference between the present Green's function and the one
used by Jack \emph{et al} and Balents and Radzihovsky is that this
decay is affected by the phonon-electron interaction (Figure
\ref{dec}b) while the process considered by them is not (Figure
\ref{dec}a). This will give a broadened peak on the current centered
at $eV = vQ$.  Physically, this is a consequence of the fact that,
once the $q=Q$ spin wave is coupled to the phonons, it is no longer a
sharp resonance.  Mathematically, we have to inspect
$G_{\mathbf{Q},\omega}$ which includes the decay of one zero-momentum
pseudospin wave with energy $\hbar\omega$ into a phonon with the same
in-plane momentum. The poles of $G_{\mathbf{Q},\omega}$ establish the
relation between the momentum $Q$ and the frequency $\omega$ and the
integral over time in equation (\ref{currentG}) gives the restriction
for the pseudospin wave energy $\hbar\omega$ to $eV$. The final
expression for the current is then
\begin{equation}\label{phoncurr}
I=-e\left(\frac{\dsas L}{4\pi\hbar\ell^2_B}\right)^2\frac{\hbar
v^2}{2\rho_s}\mathrm{Re}\frac{i(1+\lambda_D\Pi_{\mathbf{Q},\Omega})}{2\Omega(\Omega-eV/\hbar)},
\end{equation}
where $\Omega$ is the pole of $G_{\mathbf{Q},\Omega}$ with a positive
real part.

\begin{figure}
  \epsfig{file=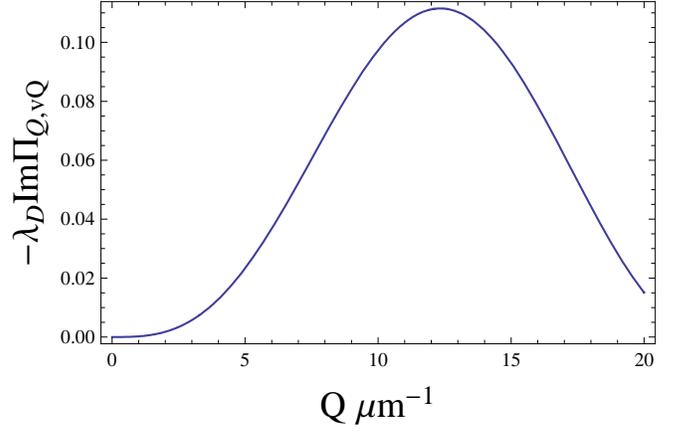, width=\columnwidth}
\caption{\small{Dimensionless coupling $\lambda_D \mathrm{Im}\Pi_{Q,vQ} $ as a function of
$Q = 2\pi B_\parallel d/\phi_0$.
($d=$28nm, $w=$18nm, $v/s=10$, $S=1$.)}}
\label{coupling:fig}
\end{figure}
A measure of the importance of the electron-phonon coupling is the
dimensionless quantity (see Fig.~\ref{coupling:fig}):
\begin{equation}
\lambda_D \mathrm{Im}\Pi_{Q,vQ} =
\frac{\lambda_DvQ}{s^3}
\sin^2\left(\frac{Qdv}{2s}\right)\exp\left(-\frac{Q^2w^2v^2}{2\pi^2s^2}\right).
\label{dimensionless:eq}
\end{equation}
This is formally small in the large-$S$ limit. Indeed, we can check
that, even if we set $S$ to unity, $\lambda_D\Pi_{Q,vQ}$ remains
small. Neglecting the geometrical form factors, the coupling strength
is given by $\lambda_D vQ/s^3 \sim 10^{-1}$ for $Q$ between 10 and 20
\inversemicron (as probed in the experiments\cite{spielman2}). The
geometrical form factors reduce this further. From
Fig.~\ref{coupling:fig}, we see that the coupling is most appreciable
for wavevectors $Q$ in the experimental range\cite{spielman2} of 10 to
20\inversemicron (with $d=28$nm and $w=18$nm). This corresponds to
$Qdv/2s\sim \pi/2$.  On the other hand, the coupling can also vanish
when $Qdv/2s$ is an integer multiple of $\pi$. At these wavevectors,
the electron and ion density oscillations are symmetric in the two
layers and so are not excited by a tunnelling event which causes
charge imbalance across the layers.

\begin{figure}[hbt]
\begin{center}
  \epsfig{file=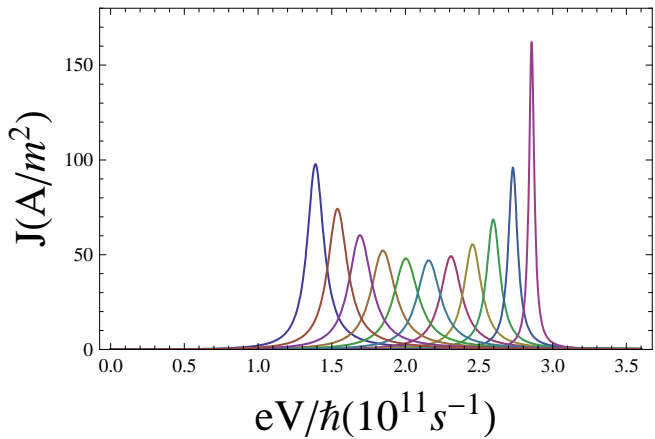, width=\columnwidth}
\caption{\small{Tunnelling current as a function of the voltage
 $eV/\hbar$ for different values of $Q= 2\pi B_\parallel d/\phi_0$
 (taken every $1\mu\mathrm{m}^{-1}$). From left to right, the first
 curve corresponds to $Q=10\mu m^{-1}$, and the last curve corresponds
 to $Q=20\mu m^{-1}$. We use $D_{\rm ef}=9$eV.}}\label{phoncurr:fig}
\end{center}
\end{figure}

The weak electron-phonon coupling means
that we can expect the pole of the Green's function to be close to the
original pseudospin wave energy: $\Omega\simeq vQ$.  We can
approximate $\Pi_{Q,\Omega}$ by $\Pi_{Q,vQ}$ and
\begin{equation}
\Omega\simeq vQ\, (1+\lambda_D\Pi_{Q,vQ}/2)\,.
\label{phonondecayrate:eq}
\end{equation}
This corresponds to a spin wave decay rate of 
$\Gamma_Q = \lambda_D vQ |{\rm Im}\Pi_{Q,vQ}|/2$.
These results give us predictions for the height and width of the
feature in the tunnelling current that disperses with the in-plane
field $B_\parallel\propto Q$ as reported by Spielman \emph{et
al}\cite{spielman2}.
From equation (\ref{phoncurr}), the maximum value of the
current occurs at $eV/\hbar=\mathrm{Re}\Omega\simeq vQ$. At
$eV/\hbar=vQ$,
\begin{equation}\label{Imax:eq}
I_\mathrm{max}
=-e\left(\frac{\dsas L}{4\pi\ell^2_B}\right)^2\frac{1}{4\hbar\rho_sQ^2}
\frac{1}{\lambda_D\mathrm{Im}\Pi_{Q,vQ}}.
\end{equation}
The width $\Delta V$ of the peak in the current as a function of the bias $V$
is similarly controlled by Im$\Omega$:
\begin{equation}
e \Delta V/\hbar = vQ \lambda_D |\mathrm{Im}\Pi_{Q,vQ}|/2\,.
\label{peakwidth:eq}
\end{equation}
First of all, the area under the peak in the $IV$ curve can be
approximated by $I_{\rm max} \Delta V$. This is proportional to $1/Q$
so that this peak weakens as we increase the in-plane field.  However,
the evolution of the shape of the $IV$ curve is a non-monotonic
function of $Q$ (see Fig.~\ref{phoncurr:fig}). The peak is sharp when
electron-phonon coupling is weak, \emph{i.e.} away from $Q \sim$
10\inversemicron. The peak is broad in the experimental regime because
of the appreciable electron-phonon coupling as we noted above.
\begin{figure}
\begin{center}
  \epsfig{file= 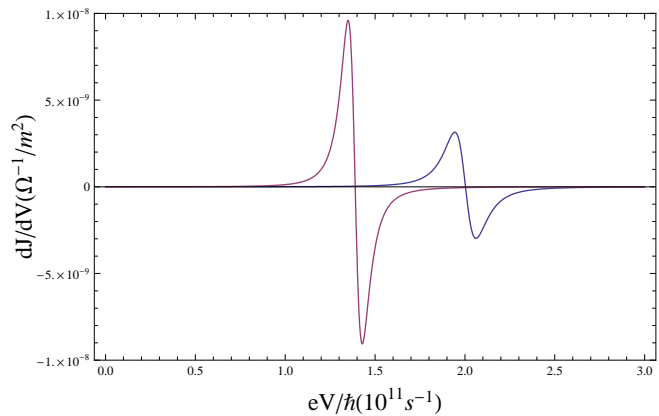, width=\columnwidth}
\caption{\small{Tunnelling density conductance
$dJ/dV$ as a function of the voltage $eV/\hbar$ for
$Q=10\mu\mathrm{m}^{-1}$ (left) and $Q=14\mu\mathrm{m}^{-1}$ (right).}}\label{dI/dV:fig}
\end{center}
\end{figure}
Our results are qualitatively consistent with the dispersive feature
in the $IV$ characteristic observed by Spielman \emph{et
al}\cite{spielman2} . This feature has been identified as due to the
excitation of coherent excitations of the interlayer-coherent phase
because its position moves linearly with the in-plane magnetic field.
The width of the feature appears to have non-monotonic dependence on
the in-plane field.  It would be interesting to explore this
dependence for a larger range of wavevectors in order to elucidate
relaxation mechanisms in the system. 

We note that our theoretical results differ in absolute magnitude from
the experiments. Our current values are nearly an order of magnitude
higher than the experimental values\cite{spielman2}. This may be due
to a strongly renormalised tunnelling amplitude at low energies due to
fluctuations.  We also predict sharper peaks than seen in experiments.
The broadest peak we obtained (which does correspond to the range of
in-plane fields studied experimentally), the theory gives a width
$\Delta V$ of the order of 10\% of the position of the peak $V = \hbar
vQ/e$. See Figure \ref{dI/dV:fig} for a direct comparison with the
differential conductance. The discrepancy may be due to other
relaxation processes which contribute additively to the broadening of
the peak. (We will discuss the contribution from elastic scattering in
the next section.)  It could also be caused by a suppression of
tunnelling due to thermal\cite{balents} and quantum\cite{jacklee}
phase fluctuations.

In summary, we have found that electron-phonon scattering may have a
measurable contribution on the broadening of the coherent feature in
the tunnelling $IV$ characteric in the quantum Hall bilayer.  In our
theory, this broadening arises from the finite lifetime of the
collective density excitation of the bilayer at wavevector $Q$ due to
the decay of electron density oscillations into lattice vibrations
with the same in-plane wavevector.  The effect is in fact strongest in
the range of in-plane magnetic fields studied experimentally, giving a
width $\Delta V$ of the order of 10\% of the position of the peak $V =
\hbar vQ/e$.

\section{Effect of disorder}\label{dis}

We now consider the effect of disorder on
the tunnelling current.  Charge inhomogeneity can be present in the
system in the case of strong disorder from the random distribution of
dopants.  This will nucleate textures in the pseudospins (merons),
producing a ground state with random vorticity in the phase of
counterflow superfluid. This has been investigated by Eastham, Cooper
and Lee\cite{easthamcooperlee09} who found suppressed tunnelling in
the sense that the length scale for counterflow current to leak across
the bilayer by tunnelling becomes enhanced by at least an order of
magnitude.

In this paper, we will focus on a system of weak disorder where the
ground state of the system is free of vortices. This applies to
relatively clean samples which have small charge variations even on
the length scale of a few magnetic lengths. We will consider in
particular spatial variations in the tunnelling amplitude across the
bilayer as a consequence of not having perfectly flat layers.
In the pseudospin model, inhomogeneous tunnelling
corresponds to inhomogeneous Zeeman coupling to the in-plane component of
the pseudospin so that the Hamiltonian is:
\begin{equation}
H_\sigma = H_0 -\dsas \sum_i \left[(1+ \sigma_i) e^{iQx_i} S_i^+ +
\mathrm{h.c.}\right]
\label{Hdisorder:eq}
\end{equation}
where $\sigma_i$ represents the fractional fluctuation around the mean
tunnelling matrix element $\dsas$. We will model the on-site
fluctuation in $\sigma_i$ as Gaussian distribution with zero mean and
variance $\sigma^2$. We will consider disorder which is short-ranged on
the scale of the magnetic length.
We can perform a perturbation theory around the zero-tunnelling ground state.
After averaging over disorder, the lowest order in perturbation theory ($\sim \dsas^2$)
would give a dc tunnelling current:
\begin{eqnarray}
I=-Se\left(\frac{\dsas L}{2\hbar l_B^2}\right)^2\!\!\!\!&&\mathrm{Re}
\int\!d^2R\int_0^\infty\!\!\!\!\!\! d\tau\, C(\mathbf{R}) \times\nonumber\\
&&(1 + \overline{\sigma_{\mathbf R} \sigma_0})
e^{-i(\mathbf{Q}\cdot\mathbf{R}+eV\tau/\hbar)}\label{currdirty:eq}
\end{eqnarray}
This is clearly not a physical result since this expression still
contains the current $I_0$ for the uniform system (see equation
(\ref{currdelta:eq})) which give a delta-function response at
$eV/\hbar = vQ$.  Therefore, we need to consider further corrections
due to the disordered tunnelling strength.  As we have learnt from the
previous section, the scattering of the pseudospin wave at wavevector
$Q$ is important to consider. In the previous section, we considered
scattering due to electron-phonon interactions. Here, we need to
consider scattering by the disorder. This process involves terms of
the form $\gamma_{\bq,\bQ}\alpha^+_{\bq} \alpha_{\bf Q}$ in the
expansion of the Hamiltonian (\ref{Hdisorder:eq}) in orders of $S$. The
matrix elements are of order $S^0$ and have the form:
\begin{equation}
	\gamma_{\bq,\bQ} = -2 \dsas \sigma_{\bq} (u_\bq +v_\bq)(u_\bQ +v_\bQ)
\label{matrixelement:eq}
\end{equation}
The decay rate of a pseudospin wave at $Q$ can be calculated from
Fermi's golden rule as $(2\pi/\hbar) \sum_\bq |\gamma_{\bq,\bQ}|^2
\delta(\epsilon_\bq - \epsilon_\bQ)$.  We see that we require
$|\bq|=Q$ since these collisions are elastic.  We can approximate
$4(u_\bq +v_\bq)^2(u_\bq +v_\bq)^2$ by $(DS/\hbar vq)(DS/\hbar vQ)$.
Averaging over disorder, we obtain a decay rate of:
\begin{eqnarray}
\Gamma_Q \simeq \frac{\dsas^2\sigma^2}{v^2\hbar^3}\frac{(DS
 l_B)^2}{\hbar vQ} = \frac{\dsas^2\sigma^2}{\hbar\rho_s Q
 l_B}\left(\frac{DS}{\rho_s}\right)^{1/2},
\label{bornapprox:eq}
\end{eqnarray}
for spatially uncorrelated disorder: $\overline{\sigma_i \sigma_j}
=\sigma^2 \delta_{ij}$.  

Analogous to our previous treatment for electron-phonon coupling with
a finite decay rate for the spin waves (\ref{phonondecayrate:eq}), 
this scattering rate due to
disorder gives a peak for the $IV$ curve of width $\Delta V =
\hbar\Gamma_Q/e$ and height:
\begin{equation}\label{Imaxdisorder:eq}
I_\mathrm{max}
=\left(\frac{\dsas L}{4\pi\ell^2_B}\right)^2\frac{ev}{8\hbar\rho_sQ}
\frac{1}{\Gamma_Q}
= \frac{L^2}{\ell_B^2}\frac{ev}{128\sigma^2 \ell_B}
\left(\frac{\rho_s}{DS}\right)^{1/2}.
\end{equation}
This gives a field-independent peak current but a peak width that decreases with increasing $Q$.

For disorder with a correlation length of
$\zeta$ with the correlation function 
$\overline{\sigma_i\sigma_j}=\sigma^2 \exp[-|{\bf r}_i-{\bf
r}_j|/\zeta]$, we obtain:
\begin{equation}
\Gamma_Q 
= \frac{\dsas^2\sigma^2}{\hbar\rho_s Q
 \ell_B}\left(\frac{DS}{\rho_s}\right)^{1/2}
 \frac{\zeta^2/\ell_B^2}{(1+Q^2\zeta^2)^{3/2}}.
\end{equation}
As a function of the in-plane field, the peak width $\Delta V$
decreases and the peak current increases as we increase the in-plane
field. This is observable if the in-plane field is large enough that
$Q\zeta$ becomes large compared to unity.

We can compare the width $\Delta V$ with the position $V= \hbar v Q/e$
of this resonant feature.  For uncorrelated disorder,
\begin{equation}
\frac{\Delta V}{V} = \left(\frac{\sigma \dsas}{\rho_s Q\ell_B}\right)^2
\end{equation}
In the experimental range, $Q\ell_B \sim 0.1$, $\dsas/\rho_s \sim 10^{-3}$
and we expect $\sigma < 1$ for the validity of the perturbation
theory. We see that this broadening is very weak. Consequently, the
peak current is apparently very large as can be seen in our expression
for the peak current (which is independent of $\dsas$).  Therefore, we
conclude that spatial fluctuations in the tunnelling amplitude does
not give rise to a strong broadening of the resonant $IV$
peak. Conversely, our results indicate that, since $\Gamma_\bQ \sim 1/Q$,
the broadening effect of this source of disorder is only observable at
much smaller values of the in-plane field.
\section{Summary}\label{sumphon:sec}

We have studied in this paper extrinsic sources of scattering for the
collective excitations ('spin waves') of the counterflow superfluid,
with the aim of understanding the broadened peak that disperses with
in-plane field $B_\parallel$ in tunnelling experiments. The $IV$
peak then reflects the spectral weight of the spin waves at the
wavevector $Q = 2\pi B_\parallel d/\phi_0$. 

We have concentrated on disorder and electron-phonon interaction as
sources of spin wave decay and hence a broadening of the $IV$
feature. Interestingly, we found that the broadening is non-monotonic
and is strongest in the experimental range of in-plane fields ($Q\sim$
10 $\mu$m$^{-1}$) because the acoustic phonons emitted after the
tunnelling event involve vibrations commensurate with the spacing
between the two quantum wells. It will be interesting to investigate
the linewidths more systematically to see if this monotonic evolution
of the lineshape can be observed in experiments.

Disorder also provides broadening. We have considered fluctuations in
the tunnelling amplitude across the bilayer. Our theory suggests that
this provides only very weak broadening. Stronger disorder would nucleate 
charged quasiparticles in the system and is beyond the scope of this paper.
This is discussed recently by Eastham \emph{et al}\cite{easthamcooperlee09}.

However, the linewidths predicted are still nearly a factor of ten smaller
than the experimental results. As mentioned above, this may be due to
phase disorder due to thermal or quantum fluctuations. It will be very
useful to have experimental measurements for a wider range of the
magnetic fields and at lower temperatures to investigate whether
electron-phonon or weak disorder are important in determining the
lineshapes in this system.

\begin{acknowledgments}
We gratefully acknowledge the financial support of EPSRC grant EPSRC-GB
EP/C546814/01.
\end{acknowledgments}


\begin{thebibliography}{99}
\bibitem{girvin} S.M. Girvin, A.H. MacDonald and J.P. Eisenstein, in
\emph{ Perspectives in Quantum Hall Effects}, ed. S. Das Sarma and
A. Pinczuk (Wiley, New York, 1997).
\bibitem{moon} K. Moon, H.~Mori, K.~Yang, S.M.~Girvin, A.H.~MacDonald, L.~Zheng, D.~Yoshioka, and S.C.~Zhang, Phys. Rev. B \textbf{51}, 5138
(1995).
\bibitem{wen} X.G. Wen and A.Zee, Phys. Rev. B \textbf{47} 2265 (1993).
\bibitem{balents}L. Balents and L. Radzihovsky, Phys. Rev. Lett.
\textbf{86}, 1825 (2001).
\bibitem{stern} A. Stern, S.M. Girvin, A.H. MacDonald, and N. Ma,
Phys. Rev. Lett. \textbf{86}, 1829(2001);
M.M. Fogler and F. Wilczek, Phys. Rev. Lett. \textbf{86}, 1833   (2001);
Y.N. Joglekar and A.H. MacDonald,Phys. Rev. Lett. \textbf{87}, 196802   (2001)
\bibitem {jacklee} R.L. Jack, D.K.K. Lee and N.R. Cooper
Phys. Rev. Lett. \textbf{93}, 126803 (2004); Phys. Rev. B \textbf{71}, 085310 (2005).
\bibitem{eis} J.P. Eisenstein, G.S. Boebinger, L.N. Pfeiffer, K.W. West
and S. He, Phys. Rev.  Lett. \textbf{68},1383 (1992); J.P. Eisenstein,
L.N. Pfeiffer, and K.W. West, Phys. Rev.  Lett. \textbf{69}, 3804
(1992).
\bibitem{mur}S.Q. Murphy, J.P. Eisenstein,
G.S. Boebinger,L.N. Pfeiffer, and K.W. West,
Phys. Rev. Lett. \textbf{72}, 728 (1994).
\bibitem{kellogg} M. Kellogg, J.P. Eisenstein, L.N.Pfeiffer, and
K.W. West, Phys. Rev. Lett. \textbf{90}, 246801 (2003).
\bibitem{spielman} I.B. Spielman, J.P. Eisenstein, L.N. Pfeiffer, and
K.W. West, Phys. Rev. Lett. \textbf{84}, 5808 (2000).
\bibitem{spielman2} I.B. Spielman, J.P. Eisenstein, L.N. Pfeiffer, and
K.W. West, 
Phys. Rev. Lett. \textbf{87}, 036803 (2001).
\bibitem{girvinleshouches}S.M. Girvin, in proceedings of Les Houches summer school on \emph{Topological Aspects of Low Dimensional Systems}, ed. A. Comtet, T. Jolicoeur, S. Ouvry, F. David (Springer-Verlag, Berlin and Les Editions de Physique, Les Ulis, 2000).
\bibitem{fertig} H.A. Fertig,
Phys. Rev. B. \textbf{40}, 1087 (1989).
\bibitem{tutuc} E. Tutuc, M. Shayegan and D.A. Huse,
Phys. Rev. Lett. \textbf{93}, 036802 (2004).
\bibitem{tiemann2} L. Tiemann, J.G.S. Lok, W. Dietsche, K. von
Klitzing, K. Muraki, D. Schuh, and W. Wegscheider, Phys. Rev. B
\textbf{77} 033306 (2008).
\bibitem{tiemann} L. Tiemann, W. Dietsche, M. Hauser, and K. von
Klitzing, New J. Phys. \textbf{10} 045018 (2008); L. Tiemann, Y. Yoon,
W. Dietsche, K. von Klitzing, and W. Wegscheider, Phys. Rev. B
\textbf{80}, 165120 (2009).
\bibitem{kun} K. Yang, K. Moon, L Zheng, A.H. MacDonald, S.M. Girvin,
D. Yoshioka, and S.C. Zhang, Phys. Rev. Lett.\textbf{72}, 732 (1994)
\bibitem{bak} P. Bak, Rep. Prog. Phys. \textbf{45} 587 (1982).
\bibitem{hanna}C.B. Hanna, A.H. MacDonald and S.M. Girvin,
Phys. Rev. B \textbf{63}, 125305 (2001).
\bibitem{mahan}G. D. Mahan, \emph{ Many-Particle Physics} (Plenum, New
York, 1990).
\bibitem{cardona} Y. Yu and M. Cardona, \emph{Fundaments of
  Semiconductors: Physics and Matherial Properties} (Springer 2001).
\bibitem{easthamcooperlee09} P.R. Eastham, N.R. Cooper and D.K.K. Lee, Phys. Rev. B
\textbf{80}, 045302 (2009).
\end{thebibliography}
\end{document}